\documentstyle[aps,prl,epsfig,psfig]{revtex}

\begin{document}

\title{Quantum Phase Transitions}
\author{M. Lavagna$^{\ast }$}
\address{Commissariat \`a l'Energie Atomique, DRFMC /SPSMS, \\
17, rue des Martyrs,\\
38054 Grenoble Cedex 9, France}

\maketitle

\begin{abstract}
We give a general introduction to quantum phase transitions in
strongly-correlated electron systems. These transitions which occur at zero
temperature when a non-thermal parameter $g$ like pressure, chemical
composition or magnetic field is tuned to a critical value are characterized
by a dynamic exponent $z$ related to the energy and length scales $\Delta $
and $\xi $.\ We show how one can derive an effective bosonic model
associated with the fluctuations of the ordering fields. Simple arguments
based on an expansion to first order in the effective interaction $u$ allow
to define an upper-critical dimension $D_{C}=4$ (where $D=d+z$ and $d$ is the
spatial dimension) below which mean-field description is no longer valid. We
present an alternative tricritical crossover approach valid at $D<D_{C}$ in
the large-$N$ limit. We emphasize the role of pertubative renormalization
group (RG) approaches and self-consistent renormalized spin fluctuation
(SCR-SF) theories to understand the quantum-classical crossover in the
vicinity of the quantum critical point with generalization to the Kondo
effect in heavy-fermion systems. Finally, we quote some recent inelastic
neutron scattering experiments performed on heavy-fermions which lead to
unusual scaling law in $\omega /T$ for the dynamical spin susceptibility
revealing critical local modes beyond the itinerant magnetism picture.We
mention new attempts to describe this local quantum critical point.
\end{abstract}

\subsubsection{Introduction}

\bigskip

The recent years have seen a renewal of interest in the study of quantum
phase transitions (QPT) motivated by experiments in cuprate superconductors,
heavy fermion compounds, organic conductors and related systems (Santa Barbara 
Proceedings 1996, Sachdev 1999). By definition a quantum phase transition is a phase
transition which occurs at zero temperature when a non-thermal parameter $g$ 
like pressure, chemical composition or magnetic field
is tuned to a critical value. The transition is characterized by quantum
fluctuations even in the ground state due to Heisenberg's uncertainty
principle. It is usually accompanied by a drastic change in the behavior of
the correlation functions.

In these lectures, we will focus on the quantum phase transitions of second
order. The associated fluctuations are characterized by two different
quantities that are respectively the energy scale $\Delta $ and the
correlation length $\xi $. In the case when the energy spectrum has a gap, $%
\Delta $ measures the difference of energy between the ground state and the
lowest excited level. If the spectrum is gapless, one can still define $%
\Delta $ as a pseudo-gap, splitting the lower-energy and the higher-energy
parts in the density fluctuation spectrum. Generally speaking $\Delta $
vanishes at the transition which occurs at a critical value $g_{c}$ of a
dimensionless parameter $g$

\begin{equation}
\Delta \sim \left| g-g_{c}\right| ^{z\nu }  \label{eq1}
\end{equation}
where $z\nu $ is a critical exponent which takes a universal value
independent of the microscopic details of the hamiltonian describing the
system.\ One will denote by $\Delta _{+}$ and $\Delta _{-}$ the values of the gap
respectively above and below $g_{c}$. Apart from an eventual constant of
proportionality, both quantities are characterized by the same critical
exponent $z\nu $.

The other scale is provided by the correlation length $\xi $. It is defined
as the length scale characterizing the exponential decay of the equal-time
correlation function $\left\langle M({\bf r},t)M({\bf 0},t)\right\rangle $
in the disordered phase at $T=0$.\ The length scale $\xi $ diverges at the
quantum critical point (QCP) defined at $T=0$ and $g=g_{c}$ according to

\begin{equation}
\xi ^{-1}\sim \left| g-g_{c}\right| ^{\nu }  \label{eq2}
\end{equation}

Putting together the vanishing of $\Delta $ and the diverging of $\xi $ at
the transition, one can write

\begin{equation}
\Delta \sim \xi ^{-z}  \label{eq3}
\end{equation}
where z is the dynamic exponent. Those critical exponents are all defined at 
$T=0$ and g$\rightarrow g_{c}$. However, they are also relevant at finite
temperature and $g\neq g_{c}$.

Figure 1 displays the schematic phase diagram$\ (T,g)$ that can be derived.
The full line is the line of phase transitions separating the long-range
ordered and the disordered phases. The dotted line corresponds to the
Ginzburg temperature $T_{G\text{ }}$below which classical fluctuations are
important. In the classical critical region, an effective classical theory
applies. The regime $I$ is the quantum disordered regime driven by quantum
fluctuations as resulting from the Heisenberg's uncertainty principle.
Oppositely only thermal fluctuations are relevant in the regime $III$ above $%
T_{G\text{ }}$.\ Finally, the regime $II$ is the quantum critical regime
characterized by a rich interplay of effects driven by both quantum and
thermal fluctuations.

\subsubsection{Experimental examples}

\bigskip

We will mention some recent experimental studies of second-order quantum
phase transitions.

\vspace{0.2in}

(i) $LiHoF_{4}$

In the insulator $LiHoF_{4}$, the $Ho$ ions fluctuate between two spin
states depending whether the spins are parallel or antiparallel to a
particular crystalline axis. At zero temperature, applying an external
magnetic field perpendicular to the magnetic axis makes the system go from a
ferromagnetic ground state to a quantum paramagnetic state in which
long-range ferromagnetic order is destroyed due to quantum tunneling between
the two spin states (Bitko et al, 1996). The quantum critical point is reached at a
critical value $H_{C}$ of the magnetic field. Raising the temperature also
destroys the long-range magnetic order driving the system to a
high-temperature magnetically disordered state. Note that the latter state,
the thermal paramagnet is fundamentally different in nature from the quantum
paramagnet formed at zero temperature.

\vspace{0.2in}

(ii) Heavy fermion compounds

One of the most striking properties of heavy fermion compounds discovered
these last years is the existence of a quantum phase transition driven by
chemical composition change, pressure or magnetic field. For example, $%
CeCu_{6-x}Au_{x}$ (von Lohneysen et al, 1994) and $Ce_{1-x}La_{x}Ru_{2}Si_{2}$ 
(Raymond et al, 1997) present a transition from a magnetically disordered phase to a
long-range antiferromagnetic order respectively at $x_{C}=0.1$ and $%
x_{C}=0.08$. We will mention other sytems as $CeIn_{3}$, $CePd_{2}Si_{2}$ 
(Mathur et al, 1998), $CeNi_{2}Ge_{2}$ (Steglich, 1996) and $U_{1-x}Y_{x}Pd_{3}$ 
(Seaman et al, 1991) exhibiting the same behavior. Once a long-range magnetic order
is set up, the effect of a pressure or a magnetic field is to drive the
system back to a magnetically disordered phase. Remarkably while far from
the quantum critical point, the magnetically disordered phase is a Fermi
liquid with large effective masses associated with the formation of heavy
quasiparticles, there are some indications that the thermal properties of
the system in the disordered phase close to the quantum critical point
differ from that of a Fermi liquid (Santa Barbara Proceedings, 1996 and Coleman, 1999). 
Typically in $CeCu_{5.9}Au_{0.1}$ (von Lohneysen et al, 1994), the specific 
heat $C$ depends on $T$
as $C/T\sim -ln(T/T_{0})$, the magnetic susceptibility as $\chi \sim
1-\alpha \sqrt{T}$, and the $T$-dependent part of the resistivity as $\Delta
\rho \sim T$ instead of $C/T\sim \chi \sim Const$ and $\Delta \rho \sim
T^{2} $ in the Fermi liquid state. The origin of this non-Fermi liquid (NFL)
behavior is presently a problem of considerable debate.

An important element in the knowledge of the quantum critical point has been
recently brought by inelastic neutron scattering (INS) experiments performed
on a $CeCu_{5.9}Au_{0.1}$ single-crystal.\ The dynamical spin susceptibility 
$\chi "({\bf q},\omega )$ near the magnetic instability wavevector ${\bf Q}$
has been found to obey an anomalous $\omega /T$ scaling law as a function of
temperature

\begin{equation}
\chi "({\bf Q},\omega )\sim T^{-\alpha }g(\omega /T)  \label{eq4}
\end{equation}
with $\alpha $ of order $0.75$.\ Moreover, such a $\omega $ and $T$
dependence stands over the entire Brillouin zone revealing in the bulk
susceptibility too. This fact strongly suggests that the spin dynamics are
critical not only at large length scales but also at atomic length scales
contrary to what happens in the traditional itinerant magnetism picture. We
will come back to this point at the end of the paper.

\bigskip

Let us focus now on the general problem of quantum phase transitions between
a Fermi liquid and a magnetically ordered phase.\ This is still a largely
controversial topic in which many aspects are not fully understood.\ Details
of the topology of the Fermi surface often matters introducing important
nesting effects and Kohn anomalies in the spin susceptibility when the
magnetic wavevector spans the Fermi surface. For reasons of simplicity, we
will not consider those effects of nesting.

\bigskip

The rest of the paper is organized as follows. In section $3$, we show how
one can integrate out the fermion fields to derive an effective bosonic
model associated with the fluctuations of the ordering fields. The resulting
effective action exhibits a dynamic exponent $z$ equal to $3$ in the
vicinity of a ferromagnetic critical point and $2$ for the antiferromagnetic
case. In section $4$, we give simple arguments based on an expansion to
first order in the effective interaction $u$ to determine an upper-critical
dimension $D_{C}=4$ (where $D=d+z$ and $d$ is the spatial dimension) below
which first order fluctuations diverge and mean-field description is no
longer valid. Section $5$ exposes an alternative tricritical crossover
approach valid at $D<D_{C}$ in the large-$N$ limit. Section $5$ gives a
short presentation of pertubative renormalization group (RG) approaches and
their links to self-consistent renormalized spin fluctuation (SCR-SF)
theories to understand the quantum-classical crossover in the vicinity of
the quantum critical point. Section $6$ generalizes the approach to the
Kondo lattice model which is believed to describe the heavy-fermion
situation Finally, we quote some recent inelastic neutron scattering
experiments performed on heavy-fermions leading to unusual scaling law in $%
\omega /T$ for the dynamical spin susceptibility which reveals critical
local modes beyond the itinerant magnetism picture inherent to pertubative
RG and SCR-SF approaches. We mention new attempts to describe this local
quantum critical point.

\subsubsection{Derivation of an effective bosonic theory}

\bigskip

We start from the Hubbard hamiltonian which represents the archetypal model
for correlated electron system. For a single nondegenerate band of electrons
of spin $1/2$, the hamiltonian is

\begin{equation}
H=\sum_{k\sigma }\varepsilon _{k}c_{k\sigma }^{\dagger }c_{k\sigma
}+U\sum_{i}n_{i\uparrow }n_{i\downarrow }  \label{eq5}
\end{equation}
where the last term the Coulomb repulsion term expressed in Wannier
representation, $\varepsilon _{k}=-[\sum_{{\bf \delta }}t_{ij}\exp (i{\bf k}.%
{\bf \delta )]/}z${\bf , }t$_{ij}$ is the overlap integrals between
neighboring sites and $z$ the coordination number. The hamiltonian is
characterized by a competition between the kinetic term and the Coulomb
interaction $U$ which generally induces a quantum phase transition from a
Fermi liquid to a magnetically ordered state at a critical value of $U/t$.

By integrating out the fermion fields, we will first show how from the
Hubbard hamiltonian, we can derive an effective bosonic theory in terms of
the fluctuations of the ordering fields $\Phi _{i\text{ }}$associated with
the magnetizations. Following (Hertz and Klenin, 1974), we
decompose the proof into the following steps. The calculations are performed
in the magnetically-disordered phase.

First let us write the Coulomb term as a function of charge and spin density
variables

\begin{equation}
U\sum_{i}n_{i\uparrow }n_{i\downarrow }=\frac{U}{4}\sum_{i}(n_{i\uparrow
}+n_{i\downarrow })^{2}-\frac{U}{4}\sum_{i}(n_{i\uparrow }-n_{i\downarrow
})^{2}
\end{equation}
where $(n_{i\uparrow }-n_{i\downarrow })/2=\Phi _{i}^{z}$ is the $z$%
-magnetization. Note that one can extend the writing to all three directions 
$x$, $y$ and $z$ in order to preserve the spin rotation invariance.

Let us consider the functional integral of the partition function

\begin{equation}
Z=\int {\cal D}c_{i\sigma }\exp \left[ -\int_{0}^{\beta }{\cal L}(\tau
)d\tau \right]  \label{eq7}
\end{equation}
\[
{\cal L}(\tau )=\sum_{i\sigma }c_{i\sigma }^{\dagger }\partial _{\tau
}c_{i\sigma }+H(\tau ) 
\]
and perform a Hubbard-Stratonovich transformation on the Coulomb interaction
term. Using the identity

\begin{equation}
\int d\Phi _{i}^{z}(\tau )\exp \left[ -\int_{0}^{\beta }(\Phi _{i}^{z}(\tau
)-\sqrt{U}c_{i\sigma _{1}}^{\dagger }\tau _{\sigma _{1}\sigma
_{2}}^{z}c_{i\sigma _{2}})(\Phi _{i}^{z}(\tau )-\sqrt{U}c_{i\sigma
_{3}}^{\dagger }\tau _{\sigma _{3}\sigma _{4}}^{z}c_{i\sigma _{4}})d\tau %
\right] =1  \label{eq8}
\end{equation}
in which the summation on the spin indices $\sigma _{i}$ are implicit and $%
\tau ^{i}$ represent the Pauli matrices, we find

\begin{equation}
Z=Z_{0}\int {\cal D}c_{i\sigma }d\Phi _{i}^{z}(\tau )\exp \left[
-\int_{0}^{\beta }(\Phi _{i}^{z}(\tau )\Phi _{i}^{z}(\tau )-\sqrt{U}\Phi
_{i}^{z}(\tau )c_{i\sigma }^{\dagger }\tau _{\sigma \sigma ^{\prime
}}^{z}c_{i\sigma ^{\prime }})d\tau \right]  \label{eq9}
\end{equation}

Integrating out the grassmannian variables $c_{i\sigma }$, we get

\begin{equation}
Z=Z_{0}\int d\Phi _{i}^{z}(\tau )\exp [-S_{eff}(\Phi _{i})]  \label{eq10}
\end{equation}

\[
S_{eff}(\Phi _{i})=\int_{0}^{\beta }\int_{0}^{\beta }\left[ \Phi
_{i}^{z}(\tau )\Phi _{i}^{z}(\tau )\delta (\tau -\tau ^{\prime })-Tr\ln
\left( 1-\sqrt{U}\Phi _{i}(\tau )G_{0}^{ij}(\tau -\tau ^{\prime })\right) %
\right] d\tau d\tau ^{\prime } 
\]
where $G_{0}$ is the bare Green function of electrons at $U=0$.

Expanding the $Trln$ term up to the second and fourth order in $\Phi $, we
have

\begin{eqnarray}
S_{eff}(\Phi _{i}) &=&\beta V\sum_{q,i\omega _{\nu }}(1-U\chi
_{_{0}}(q,i\omega _{\nu }))\Phi (q,i\omega _{\nu })\Phi (-q,-i\omega _{\nu })
\label{eq11} \\
&&+u\beta V^{4}\sum_{q_{i},i\omega _{i}}\Phi (q_{1},i\omega _{1})\Phi
(q_{2},i\omega _{2})\Phi (q_{3},i\omega _{3})\Phi
(-q_{1}-q_{2}-q_{3},-i\omega _{1}-i\omega _{2}-i\omega _{3})  \nonumber
\end{eqnarray}
where $\chi _{_{0}}(q,i\omega _{\nu })$ is the bare dynamical susceptibility
and $u$ an effective interaction assumed to be local.

Close to a ferromagnetic phase transition, we can use the Lindhard expansion
of $\chi _{_{0}}(q,\omega )$ around $q=\omega =0$

\begin{equation}
\chi _{_{0}}(q,i\omega _{\nu })=\chi _{_{0}}(0,0)-b\frac{q^{2}}{k_{F}^{2}}+ia%
\frac{\omega }{\Gamma _{q}}  \label{eq12}
\end{equation}
where $k_{F}$ is the Fermi wavevector, $a$ and $b$ are constants and $\Gamma
_{q}=qv_{F}$ is the relaxation rate which vanishes in the $q\rightarrow 0$
limit. Note that this vanishing is imposed by some symmetry arguments due to
the fact that the fluctuations of the order parameter are conserved in the
ferromagnetic case. Hence one can draw the effective bosonic action in the
ferromagnetic case

\begin{eqnarray}
S_{eff}(\Phi _{i}) &=&\beta V\sum_{q,i\omega _{\nu }}(\delta +q^{2}+\left|
\omega _{\nu }\right| /q)\Phi (q,i\omega _{\nu })\Phi (-q,-i\omega _{\nu })
\label{eq13} \\
&&+u\beta V^{4}\sum_{q_{i},i\omega _{i}}\Phi (q_{1},i\omega _{1})\Phi
(q_{2},i\omega _{2})\Phi (q_{3},i\omega _{3})\Phi
(-q_{1}-q_{2}-q_{3},-i\omega _{1}-i\omega _{2}-i\omega _{3})  \nonumber
\end{eqnarray}
where $\delta =1-U\chi _{_{0}}(q,i\omega _{\nu })$ called the Stoner factor
measures the distance to the magnetic instability.

The result can be generalized to the case of an antiferromagnetic
instability.\ In this case if we denote by $q$ the deviation to the
antiferromagnetic wavevector $Q=(\pi ,\pi ,\pi )$, the expansion expressed
in Eq. (\ref{eq12}) is still valid around $Q$. However since the
fluctuations are not conserved in the antiferromagnetic case, the relaxation
rate $\Gamma _{q}$ is now $q$-independent. The first term of Eq (\ref{eq13})
is modified in the following way

\begin{equation}
S_{eff}^{(2)}(\Phi _{i})=\beta V\sum_{q,i\omega _{\nu }}(\delta
+q^{2}+\left| \omega _{\nu }\right| )\Phi (q,i\omega _{\nu })\Phi
(-q,-i\omega _{\nu })  \label{eq14}
\end{equation}

The justification of the expansion of the $Trln$ term in Eq.(\ref{eq10}) up
to the second and fourth order in $\Phi $ lies in the fact that $\Phi _{i}$
representing the fluctuations of the magnetization are expected to be small
in the magnetically-disordered phase close to the quantum critical point.\
However, the expansion would not stand within the long-range ordered state
where the determination of the finite magnetization $M_{0}$ which $\Phi $
can be expanded\ about ($\Phi =M_{0}+\delta \Phi $) requires to minimize the
full expression of the free energy including all orders and not stop to the
fourth order. This is precisely what the equivalent of the gap equation for
the magnetization does.

\subsubsection{Pertubation theory in u : existence of an upper-critical
dimension}

In the effective bosonic theory presented in the previous section, $q$ and $%
\omega $ do not appear at the same order. They may do so in some different
models. For convenience in this section, we will only consider the case when 
$\omega $ appears at the 2nd order too and use the relastivistic notation $%
Q^{2}=q^{2}+\omega ^{2}$.\ We will show the upper-critical dimension to be
equal to $4$ for $D=d+1$ as a result of diverging critical fluctuations when 
$D<4$. The result will be generalized later on when a dynamic exponent $z$
instead of $1$ makes $D$ change to $d+z$ keeping the same value of $4$ for
the upper-critical dimension. In the realistic notations, the effective
action writes

\begin{eqnarray}
S_{eff}(\Phi _{i}) &=&\beta V\sum_{q,i\omega _{\nu }}(\delta +Q^{2})\Phi
(Q)\Phi (-Q)  \label{eq15} \\
&&+u\beta V^{4}\sum_{q_{i},i\omega _{i}}\Phi (Q_{1})\Phi (Q_{2})\Phi
(Q_{3})\Phi (-Q_{1}-Q_{2}-Q_{3})  \nonumber
\end{eqnarray}

At zero temperature and to zeroth order in $u$, the spin susceptibility $%
\chi (Q)$ related to the correlation function in $\Phi $ is given by

\begin{equation}
\lbrack \chi ^{(0)}(Q)]^{-1}=Q^{2}+\delta  \label{eq16}
\end{equation}

The uniform static susceptibility $\chi ^{(0)}(0)$ diverges at $\delta
_{c}^{(0)}=0$. $\delta $ plays the role of a tuning parameter and the
quantum phase transition takes place at $\delta _{c}^{(0)}=0$.

To first order in u, the result is changed in the following way

\begin{equation}
\lbrack \chi (Q)]^{-1}=Q^{2}+\delta +\frac{u}{2}\int \frac{d^{D}Q}{(2\pi
)^{D}}\frac{1}{Q^{2}+\delta }  \label{eq17}
\end{equation}
where $D=d+1$ and $d$ is the spatial dimension. The critical value of $%
\delta $ is given by

\begin{equation}
\delta _{c}=-\frac{u}{2}\int \frac{d^{D}Q}{(2\pi )^{D}}\frac{1}{Q^{2}+\delta
_{c}}  \label{eq18}
\end{equation}

Denoting $s=\delta -\delta _{c}$ which measures the deviation of the system
from the QCP

\begin{equation}
\lbrack \chi (Q)]^{-1}=Q^{2}+s+\frac{u}{2}\int \frac{d^{D}Q}{(2\pi )^{D}}%
\left[ \frac{1}{Q^{2}+\delta }-\frac{1}{Q^{2}}\right]  \label{eq19}
\end{equation}

For $D>4$, the integrand diverges in the $Q\rightarrow \infty $ limit and
one needs to introduce a cut-off $\Lambda $ for $Q$. One can then perform an
expansion in s

\begin{equation}
\lbrack \chi (Q)]^{-1}=Q^{2}+s\,[1-c_{1}u\Lambda ^{D-4}]  \label{eq20}
\end{equation}
where $c_{1}$ is a nonuniversal constant depending on the nature of the
cut-off. $\chi (0)$ diverges in the $s\rightarrow 0$ limit. Mean-field
critical properties still apply with small first order corrections.

For $D<4$, the integrand now converges in the $Q\rightarrow \infty $ limit.
Therefore under the condition $u\ll \Lambda ^{D-4}$, one can put $\Lambda
\rightarrow \infty $ in the integral. To first order in $u$

\begin{equation}
\lbrack \chi (Q)]^{-1}=Q^{2}+s\left[ 1-\frac{1}{2}\frac{2\Gamma \left( \frac{%
4-D}{2}\right) }{\left( D-2\right) (4\pi )^{D/2}}\frac{u}{s^{(4-D)/2}}\right]
\label{eq21}
\end{equation}

As small as $u$ is, the correction to the mean-field result is important. It
even diverges in the $s\rightarrow 0$ limit. One can then deduce an
upper-critical value for the dimension $D_{C}=4$. At $D<D_{C}$ mean-field
results are not correct and a more sophisticated resummation of the
pertubation expansion is required. For this purpose, we now present a large
N theory introducing a tricritical crossover function which allows to do
that.

\subsubsection{Large-N theory and tricritical crossovers}

\bigskip

For $D<4$, Eq. (\ref{eq21}) established at first order in $u$ suggests that $%
[\chi (Q)]^{-1}$ can be expressed as

\begin{equation}
\lbrack \chi (Q)]^{-1}=s\,\psi _{D}(q,v)  \label{eq22}
\end{equation}
where we have denoted by $q=Q/s^{1/2}$ and $v=u/s^{(4-D)/2}$ and introduced $\psi
_{D}\left[ q,v\right] $ as a universal function called the tricritical
crossover function (Br\'ezin and Zinn-Justin, 1985 and Sachdev, 1999). To first order in $u$, $\psi
_{D}(q,v)$ can be identified with

\begin{equation}
\psi _{D}(q,v)=q^{2}+1-\frac{\Gamma \left( \frac{4-D}{2}\right) }{\left(
D-2\right) (4\pi )^{D/2}}v+{\cal O}(v^{2})  \label{eq23}
\end{equation}

If we assume that we can put $\Lambda \rightarrow \infty $ in all the
higher-order terms in $u$, we can expect $[\chi (Q)]^{-1}$ to take the form
expressed in Eq. (\ref{eq22}).

We now explain how in the large $N$ limit $\psi _{D}(q,v)$ can be determined
at any order in $v$. To do that, we first extend the previous descriptions
to any value of the degeneracy of the Hubbard-Stratonovich parameter $\Phi
_{\alpha }$, $\alpha =1,2...N$ and let $N$ go to $\infty $ at the end of the
calculation. Performing a Hubbard-Stratonovich transformation on the the $%
\Phi ^{4}$-term, one gets in the large-N limit

\begin{equation}
\lbrack \chi (Q)]^{-1}=Q^{2}+\delta +u\frac{N+2}{6}\left\langle \Phi
^{2}\right\rangle  \label{eq24}
\end{equation}
where $\left\langle \Phi ^{2}\right\rangle $ is self-consistently determined
as in the self-consistent one-loop approximation according to

\begin{equation}
\left\langle \Phi ^{2}\right\rangle =\int \frac{d^{D}Q}{(2\pi )^{D}}\frac{1}{%
Q^{2}+\delta +u\frac{N+2}{6}\left\langle \Phi ^{2}\right\rangle }
\label{eq25}
\end{equation}

Hence $\delta _{C}$ is defined by

\begin{equation}
\delta _{C}+u\frac{N+2}{6}\left\langle \Phi ^{2}\right\rangle |_{\delta
=\delta _{C}}=0  \label{eq26}
\end{equation}

and 
\begin{equation}
\lbrack \chi (Q)]^{-1}=Q^{2}+(\delta -\delta _{C})+u\frac{N+2}{6}%
[\left\langle \Phi ^{2}\right\rangle -\left\langle \Phi ^{2}\right\rangle
|_{\delta =\delta _{C}}]  \label{eq27}
\end{equation}

The critical crossover function $\psi _{D}(q,v)$ can be identified with

\begin{equation}
\psi _{D}(q,v)=q^{2}+\pi _{D}(v)  \label{eq28}
\end{equation}
where the function $\pi _{D}(v)$ is solution of the following nonlinear
equation

\begin{equation}
\pi _{D}(v)+Nv\frac{\Gamma \left( \frac{4-D}{2}\right) }{3\left( D-2\right)
(4\pi )^{D/2}}\left[ \pi _{D}(v)\right] ^{(D-2)/2}=1  \label{eq29}
\end{equation}

In the $v\rightarrow \infty $ limit, $\pi _{D}(v)\sim v^{-2/(D-2)}$ and $%
[\chi (0)]^{-1}$ behaves as $s^{D/(D-2)}$ at small $s$. The latter result
on\ $[\chi (0)]^{-1}$ settles the difficulties arising from the diverging
correction term obtained to first order in u when $D<D_{C}$.

\subsubsection{\protect\bigskip Pertubative renormalization group}

\bigskip

We send the reader to a number of very good reviews existing in the
litterature on this sophistical and powerful approach. We will simply try to
clear out the essential points involved in the method. Let us start again
from the effective action derived in section 4

\begin{eqnarray}
S_{eff}(\Phi ) &=&\frac{1}{2}\sum_{q,\omega }(\delta +q^{2}+\left| \omega
\right| /q)\text{\thinspace }\left| \Phi (q,\omega )\right| ^{2}
\label{eq30} \\
&&+\frac{1}{_{2}}u\frac{1}{\beta {\cal N}}\sum_{q_{i},\omega _{i}}\Phi
(q_{1},\omega _{1})\Phi (q_{2},\omega _{2})\Phi (q_{3},\omega _{3})\Phi
(-q_{1}-q_{2}-q_{3},-\omega _{1}-\omega _{2}-\omega _{3})  \nonumber
\end{eqnarray}
where ${\cal N}$ is the number of sites and $\beta $ is the inverse
temperature. The general idea of the renormalization group approach is to
get rid of the short-range and short-time details of the fluctuations of the
order parameter to derive a renormalized effective action in which the
different parameters are rescaled. As shown in Figure 2 the aim is then to
eliminate out the contribution to $\Phi (q,\omega )$ from the outer-shell
with large values of $q$ and $\omega $. We also define the complementary
inner-shell characterized by small values of $q$ and $\omega $. Following
Hertz' paper (Hertz, 1976), we show how to derive the scaling equations along
two steps : first keeping the inner-shell contribution only, and then adding
the remaining contribution of the outer-shell. This provides us with two
scaling equations for $\delta $ and $u$, respectively the Stoner factor and
the effective interaction. We will mention how later on, Millis showed how
it is crucial to consider an additional scaling equation for temperature.

$(i)$ eliminating out the outer-shell within the inner shell contribution to 
$S_{eff}$.

First, let us remove the outer-shell contribution to $\Phi (q,\omega )$.
Since $q$ and $\omega $ appears in the quadratic term of $S_{eff}(\Phi )$ at
different orders, an anisotropic scaling procedure is required. The
outer-shell is defined by

\begin{eqnarray}
\exp (-l) &<&q<1  \label{eq31} \\
\exp (-zl) &<&\omega <1  \nonumber
\end{eqnarray}
where $l$ is infinitesimal and $z$ will be proved later on to coincide with
the dynamic exponent. Retaining the contribution to $S_{eff}(\Phi )$ of the
inner-shell only with small $q$ and $\omega $ values, one gets to the second
order

\begin{equation}
S_{eff}^{(2)}(\Phi )=\frac{1}{2}\beta {\cal N}\int_{0}^{\exp (-l)}\frac{%
d^{d}q}{(2\pi )^{d}}\int_{0}^{\exp (-zl)}\frac{d\omega }{2\pi }(\delta
^{\prime }+q^{2}+\left| \omega \right| /q)\text{\thinspace }\left| \Phi
(q,\omega )\right| ^{2}  \label{eq32}
\end{equation}
The change from $\delta $ to $\delta ^{\prime }$ is of order $l$.

Next step consists in rescaling the variables $q$ and $\omega $ in the
following way

\begin{eqnarray}
q^{\prime } &=&q\exp (l)  \label{eq33} \\
\omega ^{\prime } &=&\omega \exp (zl)  \nonumber
\end{eqnarray}

With this change of variables, we have

\begin{eqnarray}
S_{eff}^{(2)}(\Phi ) &=&\frac{1}{2}\beta {\cal N}\exp \left[ -(d+z)l\right]
\int_{0}^{1}\frac{d^{d}q^{\prime }}{(2\pi )^{d}}\int_{0}^{1}\frac{d\omega
^{\prime }}{2\pi }(\delta ^{\prime }+q^{\prime 2}\exp (-2l)+\left| \omega
^{\prime }\right| /q^{\prime }\exp \left[ -(z-1)l\right] )  \label{eq34} \\
&&\left| \Phi (q^{\prime }\exp (-l),\omega ^{\prime }\exp (-zl))\right| ^{2}
\nonumber
\end{eqnarray}

The fields $\Phi (q,\omega )$ are then rescaled so that the terms in $%
q^{\prime 2}$ and $\left| \omega ^{\prime }\right| /q^{\prime }$ in $%
S_{eff}^{(2)}(\Phi )$ are unchanged. Thanks to the anisotropic scaling
procedure introduced earlier, we are allowed to do that as soon as we choose 
$z$ equal to the dynamic exponent $3$ in the ferromagnetic case. Introducing
the scaling

\begin{equation}
\Phi ^{\prime }(q^{\prime },\omega ^{\prime })=\Phi (q,\omega )\exp \left[
-(d+z+2)l/2\right]   \label{eq35}
\end{equation}
one gets 
\begin{equation}
S_{eff}^{(2)}(\Phi )=\frac{1}{2}\beta {\cal N}\int_{0}^{1}\frac{%
d^{d}q^{\prime }}{(2\pi )^{d}}\int_{0}^{1}\frac{d\omega ^{\prime }}{2\pi }%
(\exp (2l)\delta ^{\prime }+q^{\prime 2}+\left| \omega ^{\prime }\right|
/q^{\prime })\left| \Phi (q^{\prime },\omega ^{\prime })\right| ^{2}
\label{eq36}
\end{equation}

We can see from this expression that the initial form of $S_{eff}^{(2)}(\Phi
)$ is recovered provided that $\exp (2l)\delta ^{\prime }=\delta $. This
gives us the first part of the scaling equation for $\delta $

\begin{equation}
\frac{d\delta }{dl}=2\delta  \label{eq37}
\end{equation}

In the same way, the effective action to the $4$th order in $u$ can be
expressed as

\begin{eqnarray}
S_{eff}^{(4)}(\Phi ) &=&\frac{1}{_{4}}(\beta {\cal N})^{2}u^{\prime }\exp %
\left[ -3(d+z)l\right] \exp \left[ 4(d+z+2)l/2\right] \int_{0}^{1}\frac{%
d^{d}q_{i}^{\prime }}{(2\pi )^{d}}\int_{0}^{1}\frac{d\omega _{i}^{\prime }}{%
2\pi }  \label{eq38} \\
&&\Phi (q_{1},\omega _{1})\Phi (q_{2},\omega _{2})\Phi (q_{3},\omega
_{3})\Phi (-q_{1}-q_{2}-q_{3},-\omega _{1}-\omega _{2}-\omega _{3}) 
\nonumber
\end{eqnarray}
where the summation $\prod_{i=1,2,3}$ on the subscript $i$ is implicit. In
order to keep the quartic term unchanged, one needs to transform $u$
according to

\begin{equation}
u^{\prime }\exp (\varepsilon l)=u  \label{eq39}
\end{equation}
with $\varepsilon =4-(d+z)$ enabling us to write the second scaling equation
for $u$

\begin{equation}
\frac{du}{dl}=\varepsilon u  \label{eq40}
\end{equation}

As for $\delta $, we will show in $(ii)$ that the latter scaling equation
should be laced with a second term in the righthand side to be complete.
Before ending up with $(i)$, let us point out that the scaling equation (\ref
{eq40}) provides an alternative way to define the upper-critical dimension
for $D=d+z$ generalizing to any $z$ the result previously obtained in
section 4. The spatial dimension is increased by the dynamic exponent $z$
equal to $3$ in the ferromagnetic case and to $2$ in the antiferromagnetic
case. Above the upper-critical dimension found to be $D_{C}=4$, $\varepsilon 
$ has a negative sign and $u$ is rescaled to zero. We are then left with the
quadratic term of $S_{eff}$ only and the system reaches a gaussian fixed
point. Oppositely below $D_{C}=4$, $\varepsilon $ has a positive sign. The
interaction is relevant and a non-gaussian fixed point is reached.

$(ii)$ incorporating the remaining contribution of the outer-shell to $%
S_{eff}$.

The last step consists in considering the missing contribution with large
values of $q$ and $\omega $. If one denotes by $\sum_{q,\omega }^{\prime }$ and\ $%
\sum_{q,\omega }^{"}$ respectively the summation over the inner- and the
outer-shells, the corresponding correction to $S_{eff}^{(4)}(\Phi )$ is

\begin{eqnarray}
S_{eff}^{(4)}(\Phi ) &=&\frac{u}{4\beta {\cal N}}\sum_{q,\omega }\text{ }%
^{"}\prod_{i=1}^{4}\Phi (q_{i},\omega _{i})\delta \left(
\sum_{i=1}^{4}q_{i}\right) \delta \left( \sum_{i=1}^{4}\omega _{i}\right) 
\label{eq41} \\
&&+\frac{3u}{2\beta {\cal N}}\sum_{q,\omega }\text{ }^{"}\left| \Phi
(q,\omega )\right| ^{2}\sum_{q,\omega }\text{ }^{^{\prime }}\left| \Phi
(q,\omega )\right| ^{2}  \nonumber
\end{eqnarray}

The two terms in the righthand side of the last equation corresponds to the
interaction of one particle in the outer-shell with one particle
respectively in the outer-shell and in the inner one. Part of the action
which is quadratic as $\sum_{q,\omega }$ $^{\prime }\left| \Phi (q,\omega
)\right| ^{2}$ can be resummed. The final expression for $S_{eff}$ including
all contributions from the inner- and the outer-shell up to the $4$th order
is

\begin{eqnarray}
S_{eff}^{(2)}(\Phi ) &=&\frac{1}{2}\sum_{q,\omega }\text{ }^{"}(\delta
+q^{2}+\left| \omega \right| /q)\left| \Phi (q,\omega )\right| ^{2}
\label{eq42} \\
&&+\frac{u}{4\beta {\cal N}}\sum_{q,\omega }\text{ }^{"}\prod_{i=1}^{4}\Phi
(q_{i},\omega _{i})\delta \left( \sum_{i=1}^{4}q_{i}\right) \delta \left(
\sum_{i=1}^{4}\omega _{i}\right)   \nonumber \\
&&+\frac{1}{2}\sum_{q,\omega }\text{ }^{\prime }\ln [\delta +q^{2}+\left|
\omega \right| /q+\frac{3u}{\beta N}\sum_{q,\omega }\text{ }^{"}\left| \Phi
(q,\omega )\right| ^{2}]  \nonumber
\end{eqnarray}

Expanding the $ln$ term up to the 4th order in $\Phi (q,\omega )$ enables us
to get the additional part of the scaling equations.\ Putting it together
with the truncated part of the scaling equations obtained in step $(i)$, one
can write the complete scaling equations for $\delta $ and $u$

\begin{equation}
\frac{d\delta }{dl}=2\delta +\frac{3u}{\beta {\cal N}}\sum_{q,\omega }\text{ 
}^{^{\prime }}(\delta +q^{2}+\left| \omega \right| /q)^{-1}  \label{eq43}
\end{equation}

\begin{equation}
\frac{du}{dl}=\varepsilon u-\frac{9u^{2}}{\beta {\cal N}}\sum_{q,\omega }%
\text{ }^{^{\prime }}(\delta +q^{2}+\left| \omega \right| /q)^{-2}
\label{eq44}
\end{equation}

These two equations constitute the whole set of renormalization group
equations derived by Hertz. The crossover temperature $T_{I}$ separating the
quantum to the classical regime is then defined in the following way.\ At
the end of the scaling procedure, the quantum regime is reached if only the $%
\nu =0$ term\ contributes in the Matsubara frequency sum of Eqs. (\ref
{eq43}, \ref{eq44}). Later on, Millis (Millis, 1993) corrected Hertz' paper and
showed that the temperature should be rescaled as well as $\delta $ and $u$
adding a third renormalization group equation for temperature

\begin{equation}
\frac{dT}{dl}=zT  \label{eq45}
\end{equation}

Then the criterion used to differenciate the quantum from the classical
regime is slightly different. It depends on how the temperature $T^{\ast }$
at the end of the scaling procedure compares to an energy cut-off $\omega
^{\ast }$.

We will now give a physical interpretation to $T_{I}$ in terms of the
spectrum of spin excitations as schematized in Figure 3 (). The shaded
region corresponds to the Lindhard continuum of electron-hole pair
excitations. It is defined by a finite value of the imaginary part of the
bare dynamical susceptibility. The collective mode obtained from the poles
of the inverse renormalized susceptibility is a damped magnon as reported in
Figure 3.\ In the ferromagnetic case, the dispersion of the collective mode
changes from a $q$ to a $q^{3}$ dependence at a characteristic value of the
wavevector $q^{\ast }=\xi ^{-1}\sim \sqrt{\delta }$. The energy-scale $%
\omega ^{\ast }$ corresponds to the energy of the collective mode on the
scale of the magnetic length $\xi $. The regime is either quantum or
classical depending whether $T^{\ast }$ is found to be smaller or larger
than $\omega ^{\ast }$. The two regimes are characterized by different
temperature behaviors of the physical quantities as for instance the
magnetic correlation length $\xi $ and the coefficient $\gamma $ of the
linear temperature dependence of the specific heat. Above a second crossover
temperature $T_{II}$, all the relevant modes contributing to Eqs.(\ref{eq43}%
) and (\ref{eq44}) have energies much less that $k_{B}T$ and the physical
quantities are driven by the temperature only. The results obtained by
Millis are similar to those reported in Figure 4 and only depend on the
spatial dimension $d$ and on the dynamic exponent $z$.

Note that an alternative approach to the magnetic phase transitions in the
Hubbard model has been developed in (Moriya and Kawabata, 1973) based on a self-consistent
renormalized theory of spin fluctuations (SCR-SF). It leads to results very
similar to those obtained by the pertubative renormalization group.
Comparisons between the two approaches can be found in (Millis, 1993).

\subsubsection{Heavy-Fermions and the Kondo lattice model}

\bigskip

The model that is believed to describe the physics of heavy fermions is the
Kondo lattice model where impurity spins distributed on the sites of a
periodic lattice interact with the spin of local conduction electrons via a
Kondo coupling $J_{K}$. The main features of the model result from the
competition between two energy scales (Doniach, 1977): the Kondo temperature $%
T_{K}$ corresponding to the binding energy at zero temperature of the
singlet formed by the impurity spin screened by the spin of the conduction
electrons; and the Ruderman-Kittel Kasuya-Yosida -RKKY- interactions among
neighboring spins mediated by the conduction electrons. The latter one
favors the formation of a long-range magnetic order while the former one
blocks it by Kondo screening. The consequence is the existence of a quantum
critical point at a critical value of $J_{K}$ below which a long-range
magnetic order occurs. In the case when the total number of conduction
electrons is less than one per site, all phases are in a metallic state.

\bigskip

The large $N$ expansions (Millis, 1987) which have been carried out for
these models (where $N$ simultaneously represents the degeneracy of the
conduction electrons and of the spin channels) are known to give a good
description of the Kondo effect but fail to account for the
spin-fluctuations since the RKKY interactions only appear at the order $%
1/N^{2}$. With the aim to describe the critical phenomena around the quantum
critical point of the heavy-fermion systems, it has been recently proposed 
(Lavagna and P\'epin, 1999 and 2000) a self-consistent one-loop approximation for the $%
S=1/2$ Kondo lattice model ($N=2$)\ that enlarges the standard $1/N$\
expansion theories up on the spin-fluctuation effects in complete analogy
with the self-consistent renormalized theory of spin fluctuations (SCR-SF)
developed for the Hubbard model. It results a quantum-classical crossover
at finite temperature depending whether the temperature is lower or higher
than the characteristic energy scale of the damped collective mode existing
in the vicinity of the magnetic instability.\ We refer to the paper 
(Lavagna and P\'epin, 2000) for the discussion of the related phase diagram at $d=2$ and $%
d=3 $ with the predictions of a series of crossovers in the vicinity of the
quantum critical point as reported in figure 4.

\subsubsection{Conclusion}

\bigskip

To conclude, we will say that the pertubative renormalization group (RG)
approach and the related self-consistent renormalized spin fluctuation
(SCR-SF) theory by eliminating the short-range details of the fluctuations
of the order parameter, assume that the low-energy excitations are spatially
extended and by essence do not take into account the critical local nature
of the modes at the atomic length scale. As we mentioned at the end of
section 2, there exist stong indications now based on INS experiments in
heavy-fermions, that the magnetic excitations are critically local around
the quantum critical point. This feature seems to be a characteristics of
most of the strongly-correlated electron systems including high $T_{C}$
superconductors. The existence of critical local modes is related to the
formation of local moments in the ordered phase in contrast to the itinerant
magnetism picture that is described by pertubative RG and SCR-SF theories.
Note that a scaling law in $\omega /T$ of the dynamical spin susceptibility
has been obtained theoretically in the case of spin systems as for the
random two-dimensional Heisenberg antiferromagnetic model (Sachdev, 1992).
There is an urgent need to develop theories to describe this type of local
quantum critical point in itinerant systems. In this perspective, we will
mention two recent attempts to do that : on the one hand, calculations based
on the dynamical mean field theory (Si et al, 1999) which have recently lead to
scaling law in $\omega /T$ for the dynamical spin susceptibility in
agreement with experimental observations; on the other hand supersymmetric
approaches (P\'epin and Lavagna, 1997 and P. Coleman, P\'epin, Tsvelik, 2000) 
based on a mixed fermionic-bosonic
representation of the spin.which has the advantage of capturing both the
quasiparticle and the local moment features via respectively the fermionic
and bosonic degrees of freedom.

\vspace{0.2in}

{\bf Acknowlegements :} I am grateful to Zsolt Gul\'{a}csi for giving me the opportunity to give these lectures. I would like to thank Nick Bernhoeft, Andrey Chubukov, Piers Coleman, Mucio Continentino, Jacques Flouquet, Gilbert Lonzarich, Catherine P\'{e}pin, St\'{e}phane Raymond, Louis-Pierre Regnault, Almut Schr\"{o}der and Hilbert von L\"{o}hneysen for helpful discussions.

\vspace{0.2in}

$^{\ast \text{ }}$Also Part of the Centre National de la Recherche
Scientifique (CNRS)

\vfill\eject

\centerline {\bf References}

\vspace{0.2in}

D. Bitko, T.F. Rosenbaum and G. Aeppli, Phys. Rev. Lett. 
{\bf 77}, 940 (1996)

\vspace{0.2in}

E. Br\'{e}zin and J. Zinn-Justin, Nucl. Phys. B {\bf 257},
867 (1985); \ E. Miranda, V. Dobrosavljevic and G. Kotliar, Phys. Rev. Lett. 
{\bf 78}, 290 (1997)

\vspace{0.2in}

P. Coleman, Physica B {\bf 259-261}, 353 (1999)

\vspace{0.2in}

P. Coleman, C. P\'{e}pin and A.M. Tsvelik, Phys. Rev. B 
{\bf 62}, 3852 (2000); P. Coleman, C. P\'{e}pin and A.M. Tsvelik, Nucl.
Phys. B {\bf 586}, 641 (2000)

\vspace{0.2in}

S. Doniach, Physica B {\bf 91}, 231 (1977)

\vspace{0.2in}

J.A. Hertz, Phys.Rev. B {\bf 14}, 1165 (1976)

\vspace{0.2in}

J.A. Hertz and M.A. Klenin, Phys.Rev. B {\bf 10},
1084 (1974)

\vspace{0.2in}

M.\ Lavagna and C. P\'{e}pin, Phys. Rev. B {\bf 62},
6450 (2000); M. Lavagna and C. P\'{e}pin, Acta Physica Polonica B {\bf 29},
3753 (1998)

\vspace{0.2in}

N.D. Mathur, F.M. Grosche, S.R. Julian, I.R. Walker, D.M.
Freye, R.K.W. Haselwimmer and G.G. Lonzarich, Nature {\bf 394}, 39 (1998)
and references therein

\vspace{0.2in}

A.J. Millis and P.A. Lee, Phys.Rev. B {\bf 35}, 3394
(1987); A. Auerbach and K. Levin, Phys.Rev.Lett. {\bf 57}, 877 (1986)

\vspace{0.2in}

A.J. Millis, Phys.Rev. B {\bf 48}, 7183 (1993)

\vspace{0.2in}

T.\ Moriya and A. Kawabata, J. Phys. Soc. Jpn {\bf 34},
639 (1973); T. Moriya and T. Takimoto, J. Phys. Soc. Jpn {\bf 64}, 960 (1995)

\vspace{0.2in}

C. P\'{e}pin and M.\ Lavagna, Phys. Rev. B {\bf 59},
2591 (1999)

\vspace{0.2in}

C. P\'{e}pin and M.\ Lavagna, \ Z. Phys. B {\bf 103},
259 (1997); C. P\'{e}pin and M.\ Lavagna, Phys. Rev. B {\bf 59}, 12180 (1999)

\vspace{0.2in}

S. Raymond, L.P. Regnault, S. Kambe, J.M. Mignod, P. Lejay 
and J. Flouquet, J.Low Temp.Phys. {\bf 109}, 205 (1997); S. Kambe, S. Raymond, 
L.P. Regnault, J. Flouquet, P. Lejay and P. Haen, J. Phys.Soc.Jpn {\bf 65}, 
3294 (1996)

\vspace{0.2in}

S. Sachdev and J. Ye, Phys. Rev. Lett. {\bf 69}, 2411
(1992)

\vspace{0.2in}

S. Sachdev, {\it Quantum Phase Transitions} ed. by
Cambridge University Press, Cambridge (1999)

\vspace{0.2in}

For reviews see the Proceedings of the ITP Conference in
Non-Fermi Liquid Behavior in Metals, Santa Barbara, US, 17-21 June 1996, eds
P. Coleman, B. Maple and A.J. Millis, J. Phys.: Condens. Matter {\bf 8},
no.48 (1996)

\vspace{0.2in}

A. Schr\"{o}der, G. Aeppli, E. Bucher, R. Ramazashvili
and P. Coleman, Phys. Rev. Lett. 80, 5623 (1998)

\vspace{0.2in}

A. Schr\"{o}der, G. Aeppli, R. Coldea, M. Adams, O.
Stockert, H. von L\"{o}hneysen, E. Bucher, R. Ramazashvili and P. Coleman,
Nature {\bf 407}, 351 (2000)

\vspace{0.2in}

C.L. Seaman, M.B. Maple, B.W. Lee, S. Ghamaty, M.S.
Torikachvili, J.-S. Kang, L.-Z. Liu, J. Allen and D.L. Cox, Phys.Rev.Lett. 
{\bf 67}, 2882 (1991)

\vspace{0.2in}

Q. Si, J.L.\ Smith and K. Ingersent, Int. J. Mod. Phys. B {\bf %
13}, 2331 (1999); J.L.\ Smith and Q. Si, Europhys. Lett. {\bf 45}, 228
(1999); Q. Si, S. Rabello, K. Ingersent and J.L.\ Smith, cond-mat/0011477
(2000)

\vspace{0.2in}

F. Steglich, B. Buschinger, P. Gegenwart, M. Lohmann, R.
Helfrich; C. Langhammer, P. Hellmann, L. Donnevert, S. Thomas, A. Link, C.
Geiber, M. Lang, G. Sparn and W. Assmus, J. Phys.: Condens. Matter {\bf 8},
9909 (1996)

\vspace{0.2in}

H. von L\"{o}hneysen, A. Schr\"{o}der, M. Sieck and T.
Trappmann, Phys.Rev.Lett. {\bf 72}, 3262 (1994); H. von L\"{o}hneysen,
J.Phys.Cond.Matt. {\bf 8}, 9689 (1996) and references therein

\vfill\eject

\centerline {\bf Figure captions}

\vspace{0.2in}

Figure 1 : Schematic phase diagram as a function of temperature $T$ and
dimensionless coupling $g$. The full line $T_{C}$ is the line of phase
transitions separating the long-range ordered and disordered phases. The
upper dotted line is the Ginzburg temperature. An effective classical theory
apply in the region located between the two dotted lines surrounding $T_{C}$.
Regime I is the quantum disordered regime characterized by quantum
fluctuations. Regime III is the thermally disordered regime in which all the
physical quantities are driven by temperature only. Regime II is the
intermediate quantum critical regime in which both quantum and classical
fluctuations are relevant. The arrows below the graph indicate how $g$ can
be tuned by a chemical composition change $x$ or pressure $P$.

\vspace{0.2in}

Figure 2 : Illustration on the scaling procedure in $q$ and $\omega $. The
shaded region represents the outer-shell contribution associated with the
sum $\Sigma "$ over the $q$ and $\omega $ variables. The complementary
inner-shell contribution corresponds to the sum $\Sigma ^{\prime }$ over $q$
and $\omega $.

\vspace{0.2in}

Figure 3 : Continuum of electron hole pair excitations. The dotted line
represents the damped collective mode exhibiting a change in the dispersion
at a characteristic energy $\omega ^{\ast }$ and wavevector $\xi ^{-1}\sim 
\sqrt{\delta }$. The energyscale $\omega ^{\ast }$ defines the crossover
temperature $T_{I}$ separating the quantum from the classical regime.

\vspace{0.2in}

Figure 4 : Phase diagram in the plane $(T,I)$ for dimension $d$ equal to $3$
and $z=2$ for the Kondo lattice model (Lavagna and P\'epin, 2000). The shaded region
represents the long-range antiferromagnetic phase bordered by the N\'{e}el
temperature $T_{N}$. The unshaded region marks the magnetically-disordered
regimes $I,$ $II$ and $III$ associated with different behaviors of the
system. Regime $I$ is the quantum regime in which the energy of the relevant
mode on the scale of $\xi $ is much greater than $k_{B}T$. Regime $II$ and $%
III$ are both classical regimes in which the thermal effects are important
since the fluctuations on the scale of $\xi $ have energy much smaller than $%
k_{B}T$. In Regime $II$, $\xi $ is still controlled by $(1-I)$ but the
staggered spin susceptibility is sensitive to the thermal fluctuations. In
Regime $III$, both $\xi $ and $\chi _{Q}^{^{\prime }}$ are controlled by the
temperature$.$

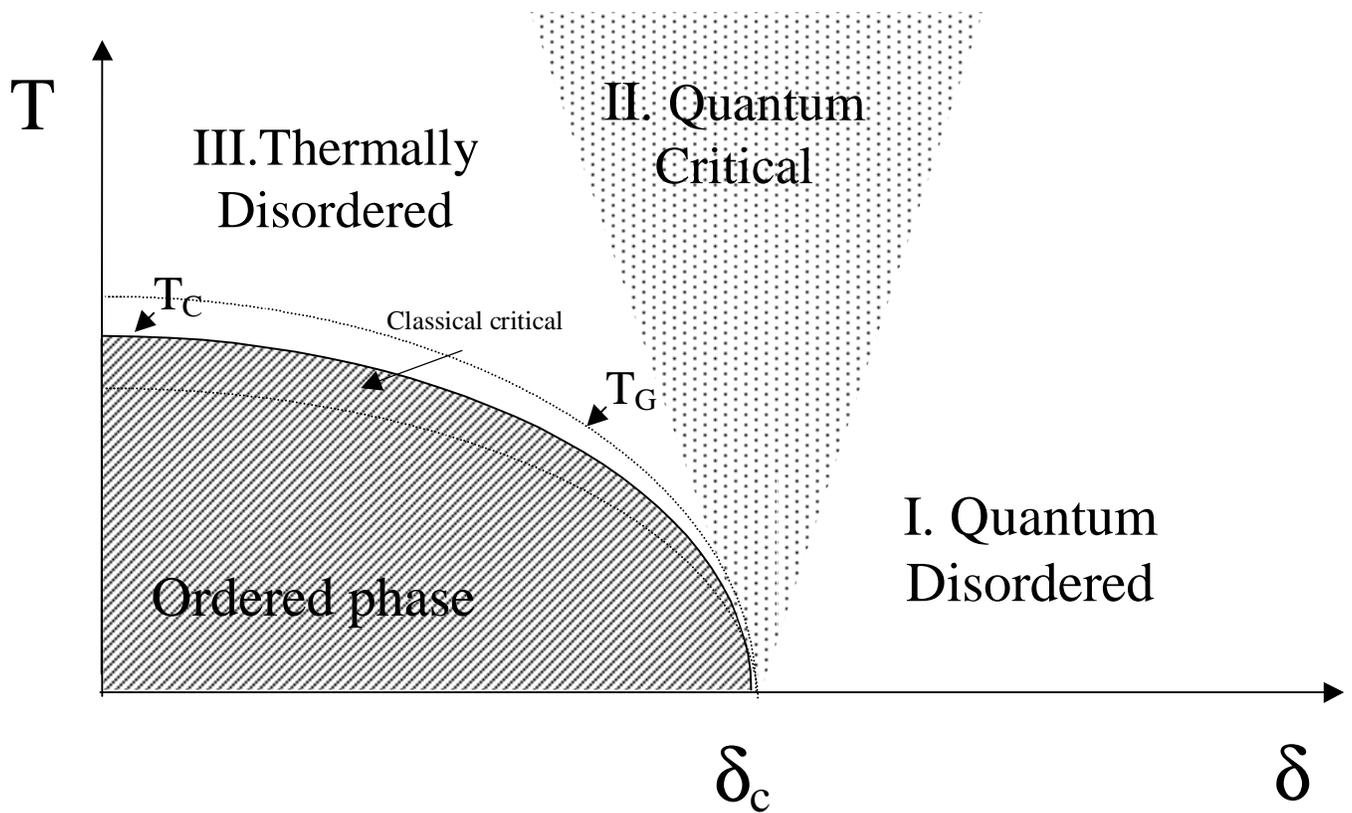
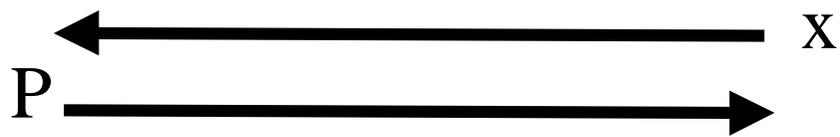
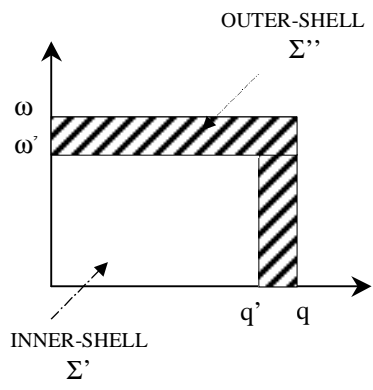
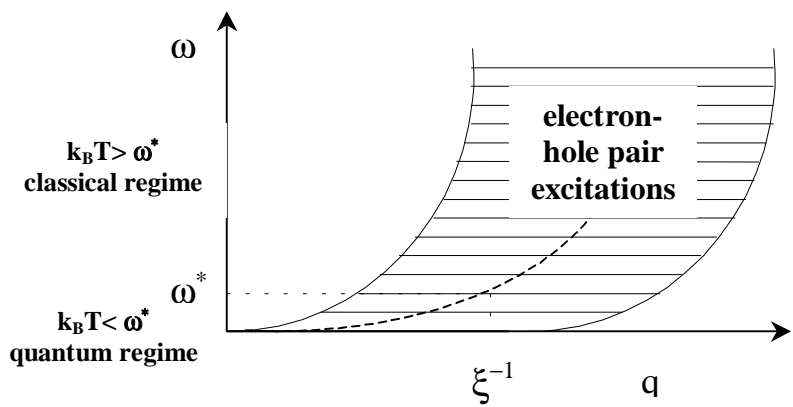
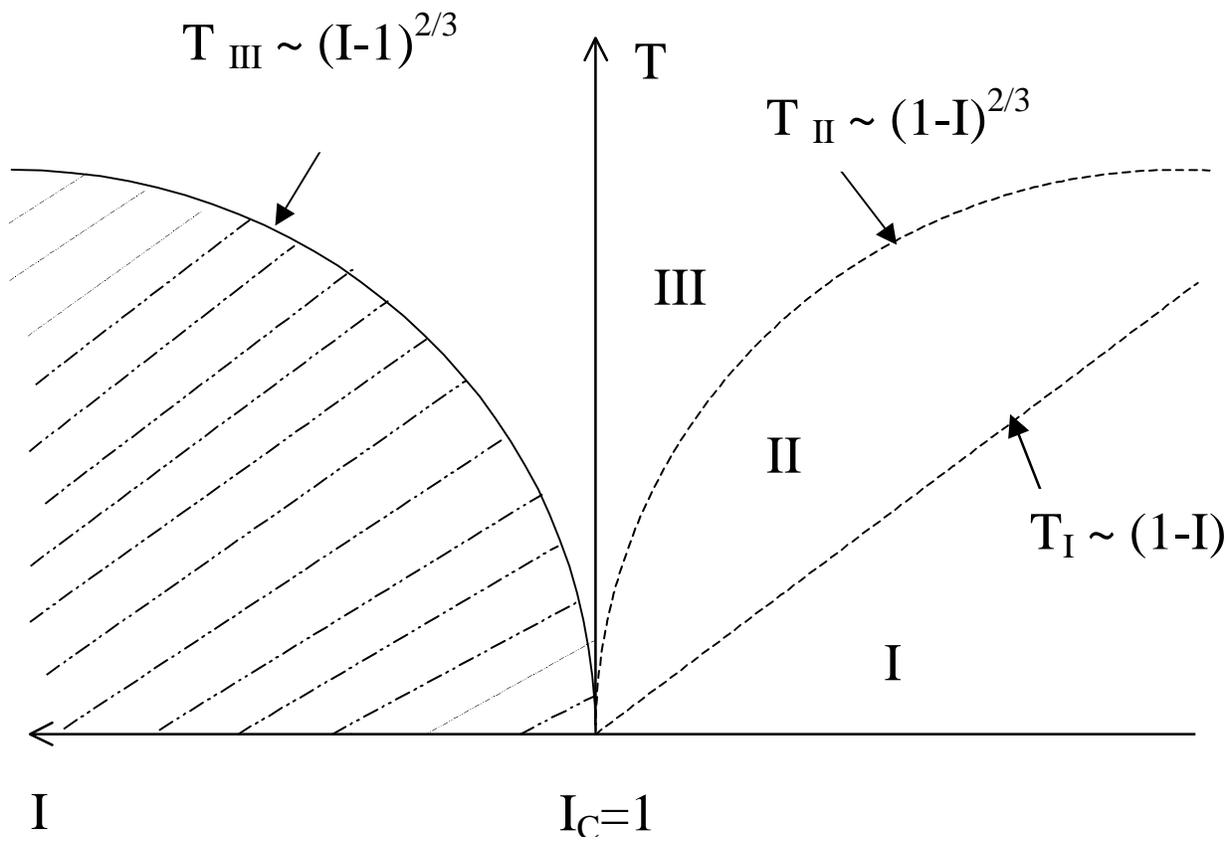
\end{document}